\def\paperID{0000}
\def\confName{CVPR}
\def\confYear{2025}

\def\paperTitle{TerraFusion: Joint Generation of Terrain Geometry and Texture\\ Using Latent Diffusion Models}

\def\authorBlock{
    Kazuki Higo 
    \qquad
     Toshiki Kanai 
     \qquad
     Yuki Endo
     \qquad
    Yoshihiro Kanamori \\
    University of Tsukuba
}

\newif\ifreview 
\newif\ifarxiv \newcommand{\arxiv}{\arxivtrue}
\newif\ifcamera 
\newif\ifrebuttal 

\arxiv 

\pdfoutput=1
\documentclass[10pt,twocolumn,letterpaper]{article}
\ifreview \usepackage[review]{cvpr} \fi
\ifarxiv \usepackage[pagenumbers]{cvpr} \fi
\ifrebuttal \usepackage[rebuttal]{cvpr} \fi
\ifcamera \usepackage{cvpr} \fi

\usepackage{graphicx}	
\usepackage{amsmath}	
\usepackage{amssymb}	
\usepackage{booktabs}
\usepackage{times}
\usepackage{microtype}
\usepackage{epsfig}
\usepackage[table,xcdraw,dvipsnames]{xcolor}
\usepackage{caption}
\usepackage{float}
\usepackage{placeins}
\usepackage{color, colortbl}
\usepackage{stfloats}
\usepackage{enumitem}
\usepackage{tabularx}
\usepackage{xstring}
\usepackage{multirow}
\usepackage{xspace}
\usepackage{url}
\usepackage{subcaption}
\usepackage{xcolor}
\usepackage[hang,flushmargin]{footmisc}

\ifcamera \usepackage[accsupp]{axessibility} \fi

\usepackage{comment}
\usepackage{booktabs}
\usepackage{multirow}
\usepackage{graphicx}
\usepackage{xfrac}
\usepackage{placeins}
\usepackage[T1]{fontenc}
\usepackage{lmodern}
\usepackage{type1cm}
\usepackage{diagbox}
\usepackage{pifont}

\usepackage{tabularx}
\usepackage{booktabs}
\usepackage[labelfont=bf,format=plain,singlelinecheck=false]{caption}
\usepackage{multirow}

\usepackage{mathcomp}

\usepackage{color}
\newcommand{\ykA}[1]{#1}
\newcommand{\higoA}[1]{#1}
\newcommand{\ykB}[1]{#1}

% ...

\ifarxiv  \fi

\newcommand{\R}[1]{{%
    \textbf{%
        \ifstrequal{#1}{1}{\textcolor{red}{R#1}}{%
        \ifstrequal{#1}{2}{\textcolor{blue}{R#1}}{%
        \ifstrequal{#1}{3}{\textcolor{magenta}{R#1}}{%
        \ifstrequal{#1}{4}{\textcolor{teal}{R#1}}{%
                           \textcolor{cyan}{R#1}%
        }}}}%
    }%
}}

\usepackage{xr-hyper}

\makeatletter
\newcommand*{\addFileDependency}[1]{
  \typeout{(#1)}
  \@addtofilelist{#1}
  \IfFileExists{#1}{}{\typeout{No file #1.}}
}

\makeatother
\newcommand*{\myexternaldocument}[1]{
    \externaldocument{#1}
    \addFileDependency{#1.tex}
    \addFileDependency{#1.aux}
}

\definecolor{cvprblue}{rgb}{0.21,0.49,0.74}
\usepackage[pagebackref,breaklinks,colorlinks,citecolor=cvprblue]{hyperref}
\usepackage[capitalize]{cleveref}
\crefname{section}{Sec.}{Secs.}
\crefname{table}{Table}{Tables}
\crefname{figure}{Fig.}{Figs.}

\ifarxiv \crefname{appendix}{App.}{Apps.}
\else \crefname{appendix}{Suppl.}{Suppls.} \fi

\unless\ifarxiv \myexternaldocument{_supplementary} \fi

\begin{document}
\title{\paperTitle}
\author{\authorBlock}
\twocolumn[{%
\renewcommand\twocolumn[1][]{#1}%
\maketitle
\includegraphics[width=1\linewidth]{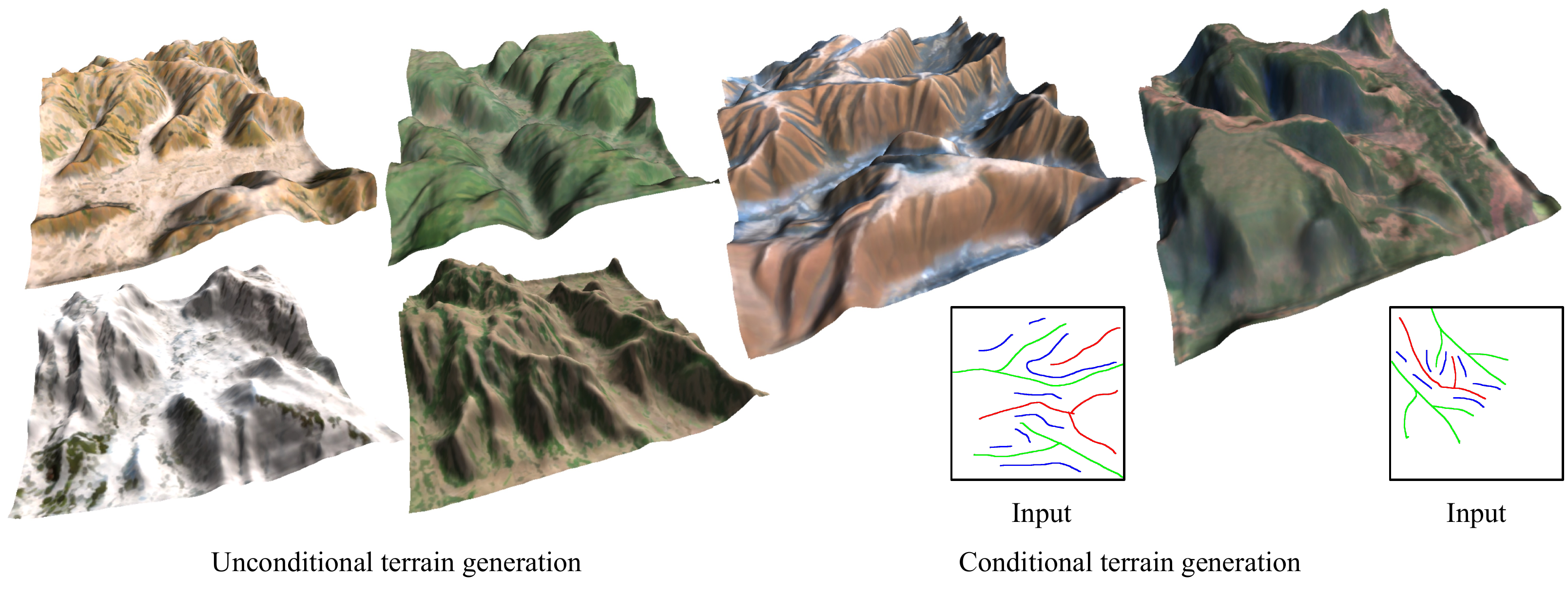}
\captionof{figure}{We introduce \textit{TerraFusion}, a novel diffusion-based framework for jointly generating terrain geometry and textures. Our method not only enables the random synthesis of plausible terrain models (left), but also supports intuitive user control through rough sketches (right), where valleys, ridgelines, and cliffs are indicated by red, green, and blue lines, respectively.\vspace{2em}}
\label{fig:teaser}
}]
\begin{abstract}
3D terrain models are essential in fields such as video game development and film production. Since surface color often correlates with terrain geometry, capturing this relationship is crucial to achieving realism. However, most existing methods generate either a heightmap or a texture, without sufficiently accounting for the inherent correlation.
In this paper, we propose a method that jointly generates terrain heightmaps and textures using a latent diffusion model. First, we train the model in an unsupervised manner to randomly generate paired heightmaps and textures. Then, we perform supervised learning of an external adapter to enable user control via hand-drawn sketches.
Experiments show that our approach allows intuitive terrain generation while preserving the correlation between heightmaps and textures.
\end{abstract}
\section{Introduction}
3D terrain models play a crucial role in creating realistic landscapes and enhancing user experiences in various applications, including video games and film production. However, manually designing realistic terrain is challenging due to the inherent correlation between terrain geometry and texture. For example, steep regions often exhibit rocky or forested textures, with color and pattern variations aligned with the terrain's contours.

Numerous methods have been proposed for automatic terrain generation~\cite{DBLP:journals/cgf/GalinGPCCBG19}.
Recent advances in deep learning have particularly enabled efficient and high-quality terrain generation~\cite{guerin2017interactive, zhang2022authoring, lochner2023interactive, DBLP:journals/cgf/PerchePBGG23, kanai2024seasonal}.
These approaches typically represent terrain geometry using heightmaps that store elevation values per pixel, framing terrain generation as an image generation task.
Examples include generating heightmaps using Generative Adversarial Networks (GANs)~\cite{guerin2017interactive,zhang2022authoring} and diffusion models~\cite{lochner2023interactive}.
There is also a method that generates textures with seasonal variation conditioned on heightmaps~\cite{kanai2024seasonal}.
\ykA{These existing methods can synthesize either heightmaps or textures. However, consistent synthesis and editing of both heightmaps and textures are not straightforward by combining these methods.}

One possible solution is a two-stage approach that combines a heightmap generator with a texture generator.
In this setup, the first stage generates \ykA{a} heightmap, and the second stage generates \ykA{a} texture conditioned on it\ykA{, and vice versa}.
However, such \ykA{two-stage} approaches are prone to error accumulation across stages, which can degrade output quality.
Another challenge is \ykA{to preserve} the correlation between the heightmap and texture during user editing.
For instance, if a user modifies the second-stage output via a sketch, the correlation with the first-stage output may be lost.

To address these issues, we propose \textit{TerraFusion}, a direct approach that models the joint distribution of heightmaps and textures (see Figure~\ref{fig:teaser}).
Specifically, we employ a Latent Diffusion Model (LDM)~\cite{rombach2022high} to generate both components simultaneously.
We extend an LDM to handle paired heightmaps and textures, and train the model in an unsupervised manner using real-world terrain data.
To reduce computational cost, the diffusion model operates in a low-dimensional latent space learned by a Variational Autoencoder (VAE).
We train a VAE \ykA{dedicated to} heightmaps to ensure accurate latent representation.
To improve output quality, we fine-tune the model using prior knowledge from Stable Diffusion~\cite{rombach2022high}, which was pre-trained on large-scale image datasets.
This joint approach avoids error accumulation inherent in sequential models and preserves the correlation between geometry and color during user editing.
Prior work has shown that diffusion models are applicable to various image editing tasks~\cite{DBLP:journals/corr/abs-2402-17525}; as one such application, we incorporate a conditioning module into the LDM to enable user control via sketch input.

Our key contributions are summarized as follows:
\begin{enumerate}
    \item \textbf{Joint generation framework}: We propose TerraFusion, a novel framework based on an LDM that simultaneously generates terrain heightmaps and textures in a fused latent space.
    \item \textbf{Domain-specific VAE}: We train a VAE \ykA{dedicated to} heightmaps, achieving more accurate reconstructions than generic image-based VAEs.
    \item \textbf{User-guided control}: We incorporate a conditioning module into our architecture, enabling intuitive and flexible terrain design from input sketches.
\end{enumerate}
Compared to baselines, our approach produces better qualitative and quantitative results and more accurately captures the correlation between heightmaps and textures.
Project page: \href{https://millennium-nova.github.io/terra-fusion-page/}{\color{magenta}{https://millennium-nova.github.io/terra-fusion-page/}}.
\section{Related Work}
\label{sec:related-work}

\paragraph{
Geometry generation.}
Classical terrain generation methods include procedural and physics-based approaches~\cite{galin2019survey}.
Procedural methods generate terrain geometry automatically using random noise and fractals, while physics-based methods simulate natural processes such as erosion and sedimentation.
However, both approaches typically require expert knowledge and iterative parameter tuning.

To make terrain generation more accessible, deep learning-based methods have been explored.
Guérin et al.~\cite{guerin2017interactive} proposed a framework based on conditional Generative Adversarial Networks (cGANs) trained on real-world terrain data.
Their system enables users to generate terrain models by sketching features such as ridgelines, rivers, and elevation reference points.
Zhang et al.~\cite{zhang2022authoring} extended this approach to support control over both elevation and terrain style.
Lochner et al.~\cite{lochner2023interactive} employed diffusion models to generate higher-quality heightmaps, also allowing users to control terrain geometry and style through sketches.
However, these methods focus exclusively on generating heightmaps and do not address texture generation.

\paragraph{
Texture generation.}
Zhu et al.~\cite{zhu2021seamless} proposed a method for generating satellite imagery from labeled map data.
Dachsbacher et al.~\cite{dachsbacher2006procedural} reconstructed terrain textures from satellite images using heightmaps, temperature, precipitation, and solar radiation data.
Kanai et al.~\cite{kanai2024seasonal} introduced a method for generating terrain textures that reflect seasonal variations.
However, these approaches focus solely on texture generation and do not address heightmap generation.

\paragraph{
Joint generation. 
}
Spick et al.~\cite{spick2019realistic} proposed a method for jointly generating heightmaps and textures using GANs.
They trained a Spatial GAN~\cite{zhao2020spatialgan} on a satellite image dataset containing both terrain heightmaps and corresponding textures, representing them as four-channel images to enable joint generation.
However, this method is limited to random terrain generation and does not support user control.
Moreover, GANs typically produce lower-quality outputs than more recent diffusion-based methods and offer limited flexibility for image editing.
In contrast, our method leverages diffusion models to achieve high-quality terrain generation with user-guided control.

Concurrently, MESA~\cite{mesa2025} explores text-driven terrain generation using an LDM trained on global terrain data. In contrast to our approach, which focuses on localized and sketch-based control, MESA operates at a global scale with textual conditioning. While MESA employs separate heads for heightmaps and textures, our method jointly generates two modalities within a fused latent space, 
supported by a VAE tailored for elevation data. 
Moreover, MESA provides only qualitative results of its method, whereas we conduct extensive evaluations, both quantitative and qualitative, to demonstrate the effectiveness of our method over various baselines.
\section{
Background
}
\label{sec:diffusion-models}
\paragraph{
Latent Diffusion Model (LDM). 
}
LDM~\cite{rombach2022high} is an image generation model based on diffusion models~\cite{ho2020denoising}.
To improve computational efficiency, it uses a VAE~\cite{Kingma2014auto} to compress input images into a low-dimensional latent space, where the diffusion process is applied during training and inference.
The compression and reconstruction of an image $\mathbf{x} \in \mathbb{R}^{H \times W \times \ykA{C}}$ \ykA{(where $H$, $W$, and $C$ are the height, width, and number of channels of the image, respectively)} using the VAE encoder $\mathcal{E}$ and decoder $\mathcal{D}$ are defined as follows:
\begin{align}
\mathbf{z} = \mathcal{E}(\mathbf{x}), \quad \hat{\mathbf{x}} = \mathcal{D}(\mathbf{z}),
\label{eq:ldm-encoder}
\end{align}
where $\mathbf{z} \in \mathbb{R}^{h \times w \times c}$ denotes the low-dimensional latent representation \ykA{and $h$, $w$, and $c$ are the height, width, and number of channels of the latent, respectively}. The diffusion process in this latent space is given by:
\begin{align}
  \mathbf{z}_t = \sqrt{\bar{\alpha}_t} \mathbf{z}_0 + \sqrt{1 - \bar{\alpha}_t} \epsilon, \quad \epsilon \sim \mathcal{N}(\mathbf{0}, \mathbf{I}),
  \label{eq:ddpm-sample-x_t}
\end{align}
where $t$ is the timestep, $\bar{\alpha}_t$ is a constant, and $\mathbf{I}$ is the identity matrix.
In the reverse diffusion process, a neural network $\epsilon_\theta$ is trained to predict the added noise by minimizing the following loss:
\begin{align}
\ykA{\mathcal{L}}_{\mathrm{LDM}} := \mathbb{E}_{\mathbf{z}, \epsilon \sim \mathcal{N}(\mathbf{0}, \mathbf{I}), t} \left[ | \epsilon - \epsilon_\theta(\mathbf{z}_t, t) |_2^2 \right].
\label{eq:ldm-loss}
\end{align}
To generate an image, the process starts with a noise vector $\mathbf{z}_T \sim \mathcal{N}(\mathbf{0}, \mathbf{I})$.
Noise is then iteratively predicted and removed using $\epsilon_\theta$, and the final latent $\mathbf{z}_0$ is decoded by $\mathcal{D}$ to produce the output image.
In this paper, we extend the LDM framework to jointly generate heightmaps and textures, and to support user-guided control.

\section{
Method
}
\begin{figure}[t]
  \centering
  \includegraphics[width=\linewidth]{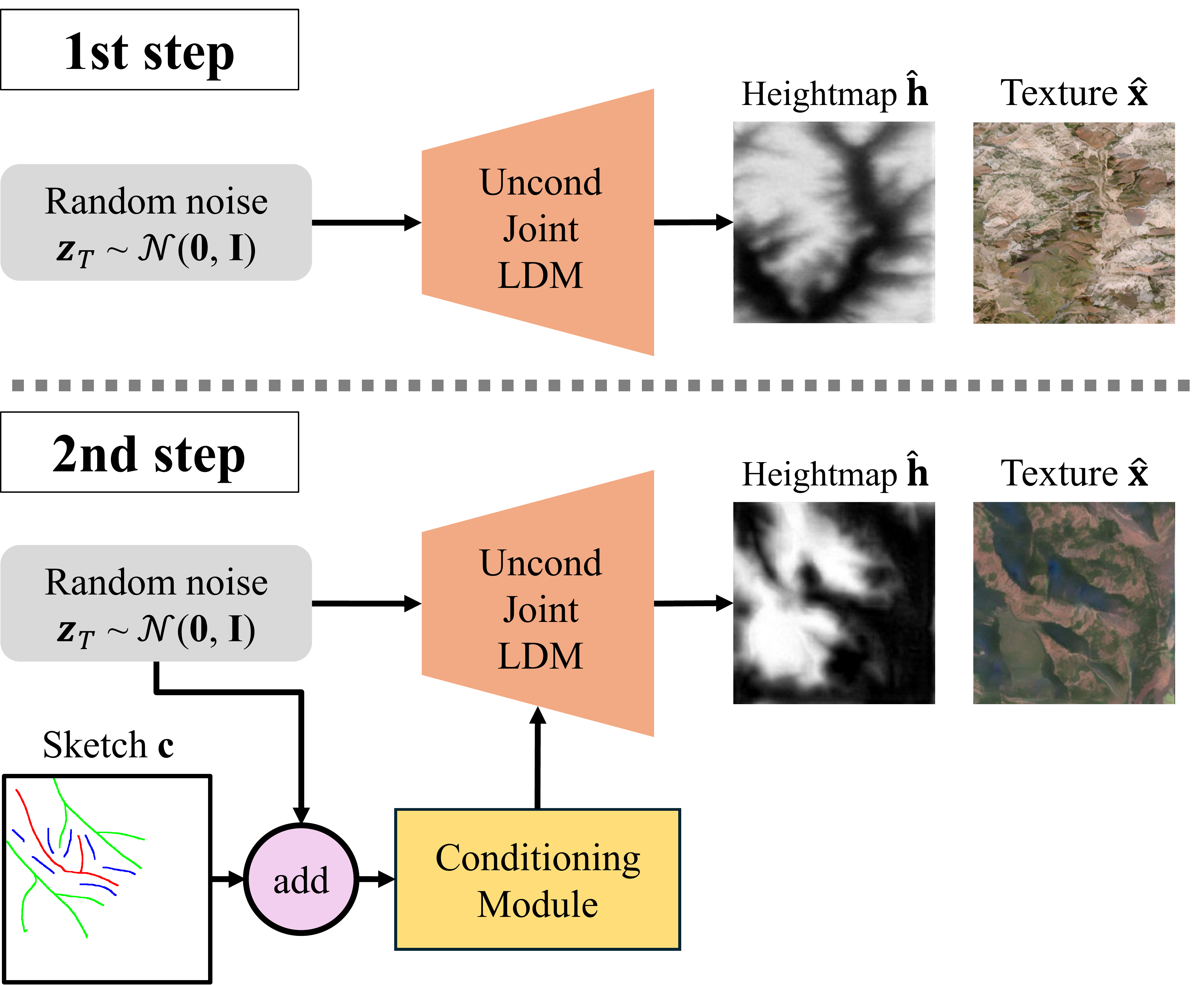}
  \caption{
  Overview of our method. Our approach consists of two stages: (1) training \ykA{of} an unconditional joint generative model that randomly generates heightmaps and textures, and (2) training \ykA{of} a conditioning module that enables output control based on inputs such as user sketches.
  }
  \label{fig:overview}
\end{figure}

An overview of our method is shown in Figure~\ref{fig:overview}.
Our approach builds the terrain generation model in two stages.
In the first stage, we train an LDM~\cite{rombach2022high} in an unsupervised manner to model the joint distribution $p(\mathbf{h}, \mathbf{x})$, where $\mathbf{h} \in \mathbb{R}^{H \times W}$ is a heightmap and $\mathbf{x} \in \mathbb{R}^{H \times W \times 3}$ is the corresponding texture.
In the second stage, we incorporate user input such as \ykA{sketches}, represented as a conditioning signal $\mathbf{c} \in \mathbb{R}^{H \times W \times 3}$, and extend the model to learn the conditional distribution $p(\mathbf{h}, \mathbf{x} \mid \mathbf{c})$.
We achieve this by adopting ControlNet~\cite{zhang2023adding} and training an external adapter in a supervised manner to condition the generation process on $\mathbf{c}$.

\subsection{
Dataset}
\label{sec:dataset}
We used NASADEM elevation data\footnote{https://lpdaac.usgs.gov/products/nasadem\_hgtv001/} (30~m spatial resolution) for heightmaps, and atmospherically corrected Sentinel-2 imagery\footnote{https://sentiwiki.copernicus.eu/web/s2-processing\#S2Processing-CopernicusSentinel-2Collection-1AvailabilityStatus} (Level-2A product, 10~m spatial resolution) for terrain textures.
We focused on natural terrain between 60\textdegree S and 60\textdegree N, extracting $1\tcdegree \times 1\tcdegree$ geographic regions from 276 land areas (see Appendix~\ref{appendix_dataset} for \ykA{the} details).
Each region was resampled to a spatial resolution of 25~m, producing images of approximately $4096 \times 4096$ pixels.
These were divided into non-overlapping $256 \times 256$ patches, 
excluding \ykA{data with missing texture details due to clouds and shadows}.
To match the input size required by the diffusion model, each patch was then upsampled to $512 \times 512$. 
The elevation values in the heightmaps ranged from 0 to 8000~m.
However, as discussed in Section~\ref{sec:heightmap-vae}, VAEs struggle to capture this wide distribution accurately.
To mitigate this, we excluded \ykA{minor} samples with elevations above 2000~m.
This preprocessing yielded a total of 4,119 paired samples of heightmaps and texture images.

\subsection{
VAE Training}
\label{sec:heightmap-vae}
\begin{figure}[t]
  \centering
  \includegraphics[width=1.\linewidth]{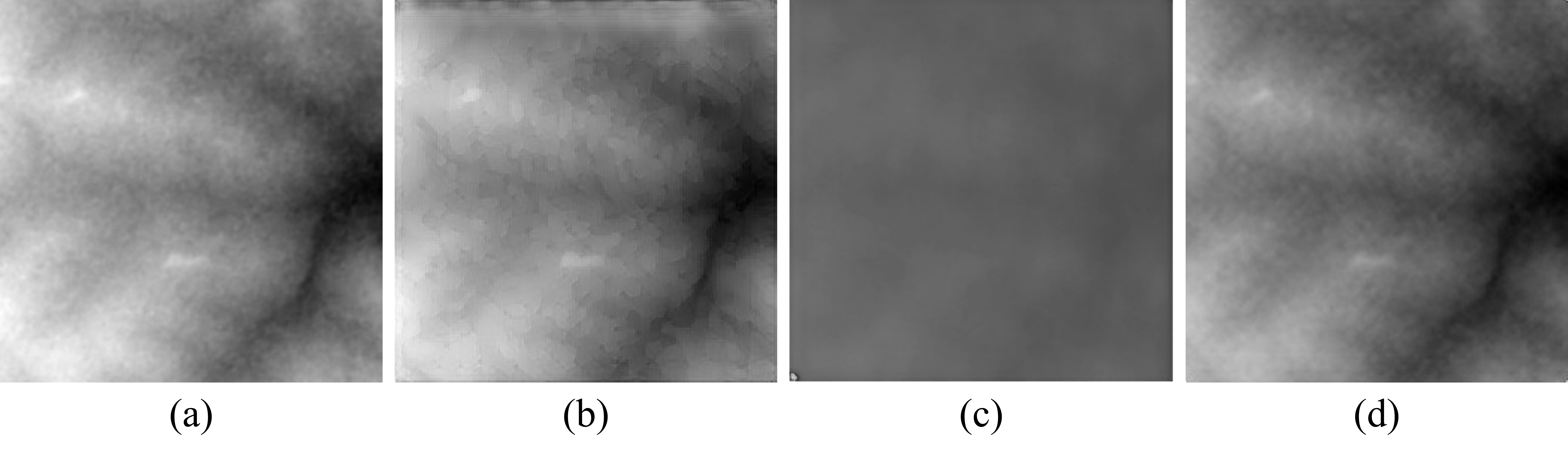}
  \caption{
  Comparison of heightmap reconstruction results. (a) Input image, (b) Reconstruction using the VAE from Stable Diffusion, (c) Reconstruction using a VAE trained on data with a maximum elevation of 8000~m, (d) Reconstruction using a VAE trained on data with a maximum elevation of 2000~m. Zooming in is recommended to better observe fine-grained differences.
  }
  \label{fig:heightmap-vae}
\end{figure}

For texture images, we use the VAE from Stable Diffusion~\cite{rombach2022high}.
However, applying this VAE to heightmaps leads to poor reconstruction quality due to a mismatch between the VAE's training distribution and elevation distribution (see Figure~\ref{fig:heightmap-vae} (b)).
To resolve this, we train a VAE specifically for heightmaps.
We adopt a Kullback-Leibler (KL) Autoencoder with the same architectural design as the VAE used in Stable Diffusion.

\paragraph{
Normalization.}
It is \ykA{a} standard practice to normalize the inputs and outputs of a VAE to the range $[-1, 1]$ for compatibility with activation functions.
Accordingly, we normalize heightmaps in this study using the following equation:
\begin{align}
  \mathbf{h}_{scaled} = \left(\frac{\mathbf{h}}{H_\text{max}} - 0.5\right) \times 2,
  \label{eq:heightmap-scaling}
\end{align}
where $\mathbf{h}$ is the original heightmap, $\mathbf{h}_{\text{scaled}}$ is the normalized version, and $H_{\text{max}}$ denotes the maximum elevation value in the dataset.

\begin{figure}[t]
  \centering
  \includegraphics[width=\linewidth]{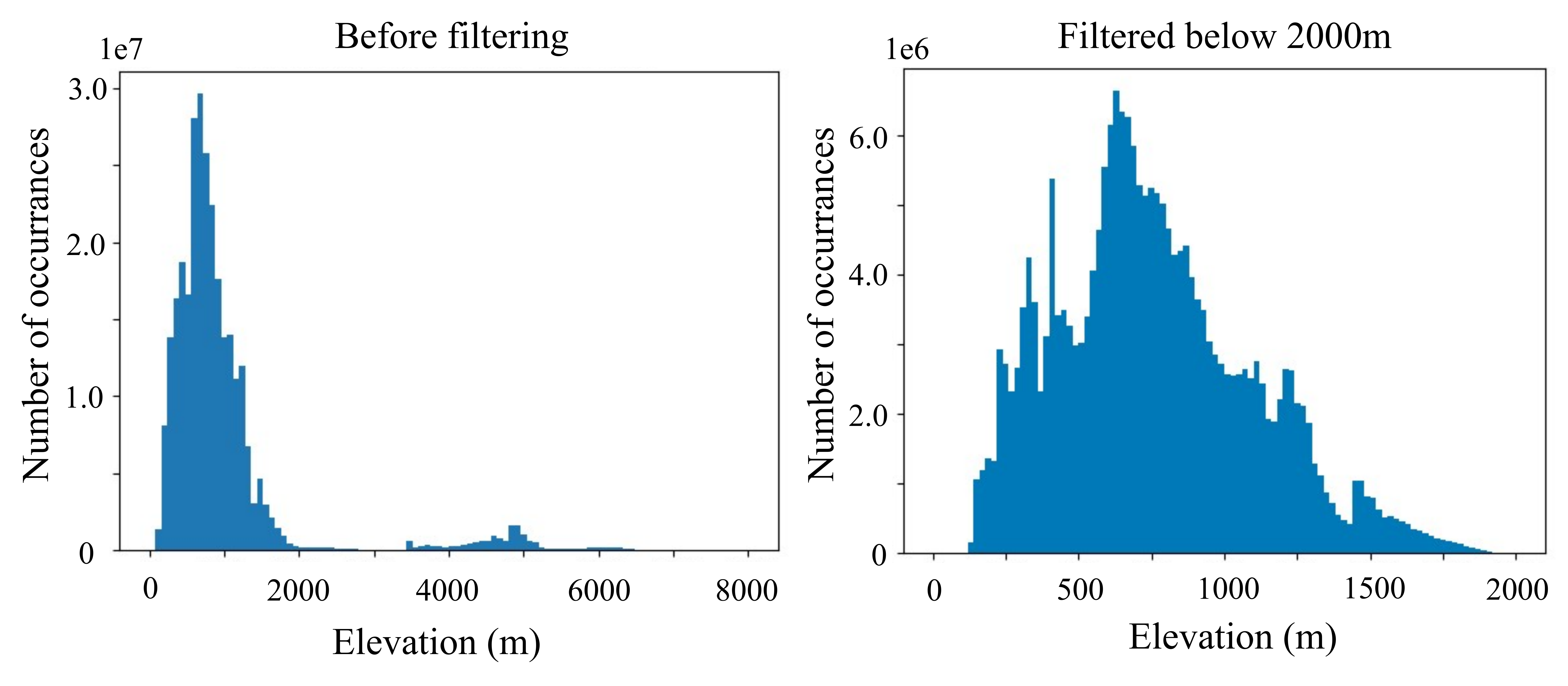}
  \caption{
  Elevation distribution of the heightmap dataset before (left) and after (right) filtering. 
  }
  \label{fig:heightmap-distribution}
\end{figure}

As mentioned in Section~\ref{sec:dataset}, although elevation values from NASADEM range from 0 to 8000~m, the VAE struggles to capture elevation variation when using $H_\text{max} = 8000$ for normalization in Equation~(\ref{eq:heightmap-scaling}). This normalization results in poor reconstruction quality (see Figure~\ref{fig:heightmap-vae} (c)). 
As shown in Figure~\ref{fig:heightmap-distribution}, most heightmaps in the dataset have elevations below 2000~m.
To improve reconstruction quality, we excluded heightmaps with elevations above 2000~m and normalized the remaining data using $H_\text{max} = 2000$.
This normalization yields \ykA{better} reconstruction results (see Figure~\ref{fig:heightmap-vae} (d)).
Handling higher-elevation terrain is left as future work.

\subsection{
Unconditional Joint Generative Model
}
\label{sec:uncond-terrain-ldm}

\begin{figure}[t]
  \centering
  \includegraphics[width=\linewidth]{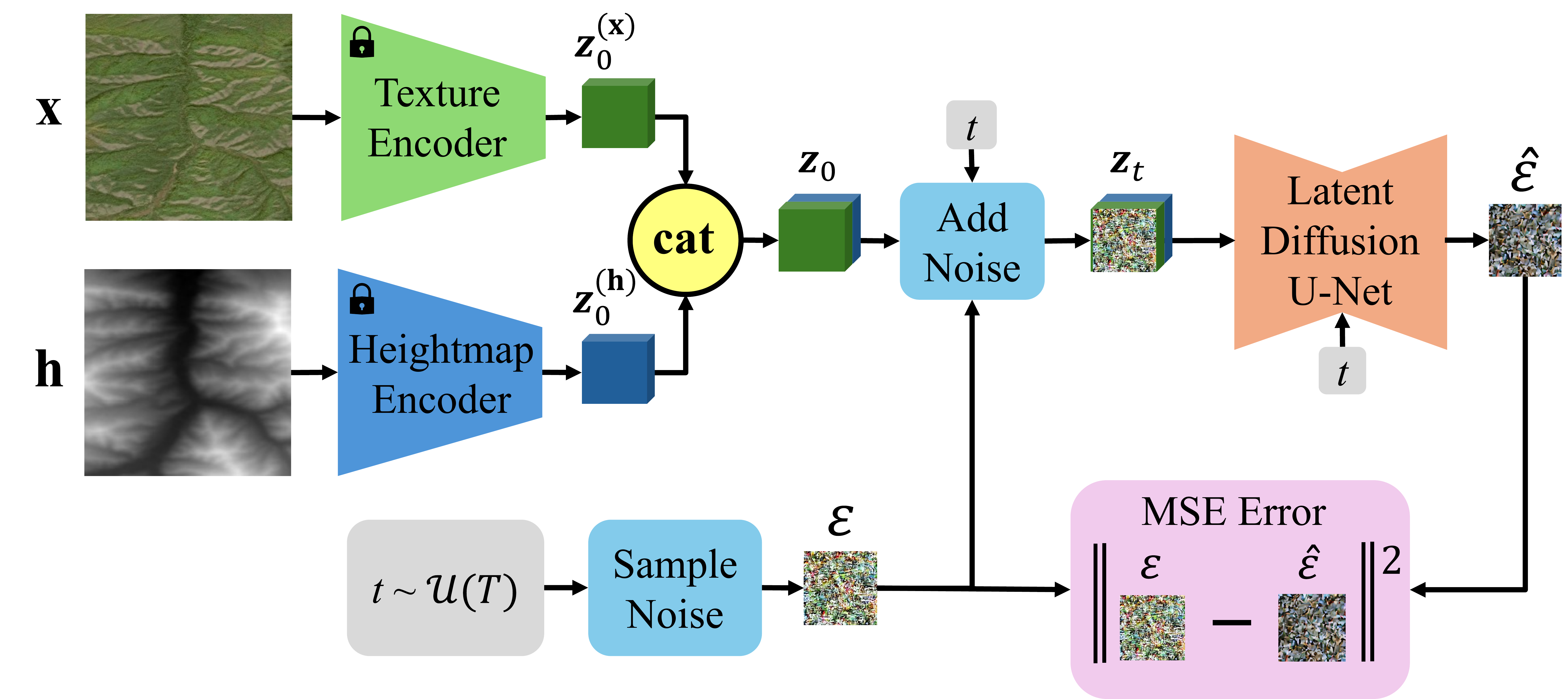}
  \caption{
  Training flow of the unconditional joint generative model.
  }
  \label{fig:train-uncond-terrain-ldm}
\end{figure}

\paragraph{Training.}
We train an LDM to jointly generate heightmaps and texture images in the latent space of the VAE introduced in Section~\ref{sec:heightmap-vae}.
Figure~\ref{fig:train-uncond-terrain-ldm} illustrates the training architecture.
Given a heightmap $\mathbf{h}$ and a texture image $\mathbf{x}$, they are encoded into their respective latent representations $\mathbf{z}^{(\mathbf{h})}$ and $\mathbf{z}^{(\mathbf{x})}$ as follows:
\begin{align}
  \mathbf{z}^{(\mathbf{h})} = \mathcal{E}_{h}(\mathbf{h}), \quad \mathbf{z}^{(\mathbf{x})} = \mathcal{E}_{x}(\mathbf{x}),
  \label{eq:terrain-ldm-encoder}
\end{align}
where $\mathcal{E}_h$ is the encoder for heightmaps, and $\ykA{\mathcal{E}_x}$ is the encoder from Stable Diffusion.
Noise is added to these latent representations using the diffusion process defined in Equation~(\ref{eq:ddpm-sample-x_t}), and a U-Net $\epsilon_\theta$ is trained to predict the added noise from the noisy inputs.
The loss function $\ykA{\mathcal{L}}$ is defined as follows:
\begin{align}
\ykA{\mathcal{L}} := \mathbb{E}_{\mathbf{z}^{(\mathbf{h})}_0, \mathbf{z}^{(\mathbf{x})}_0, \mathbf{\epsilon} \sim \mathcal{N}(\mathbf{0}, \mathbf{I}), t} \left[ \left| \mathbf{\epsilon} - \mathbf{\epsilon}_\theta(\mathbf{z}^{(\mathbf{h})}_t, \mathbf{z}^{(\mathbf{x})}_t, t) \right|^2 \right].
\label{eq:terrain-ldm-loss}
\end{align}
While the original U-Net predicts noise for a single latent variable $\mathbf{z}_t$,
our method predicts noise for both the heightmap and texture latents, $\mathbf{z}^{(\mathbf{h})}_t$ and $\mathbf{z}^{(\mathbf{x})}_t$, respectively.

\begin{figure}[t]
  \centering
  \includegraphics[width=\linewidth]{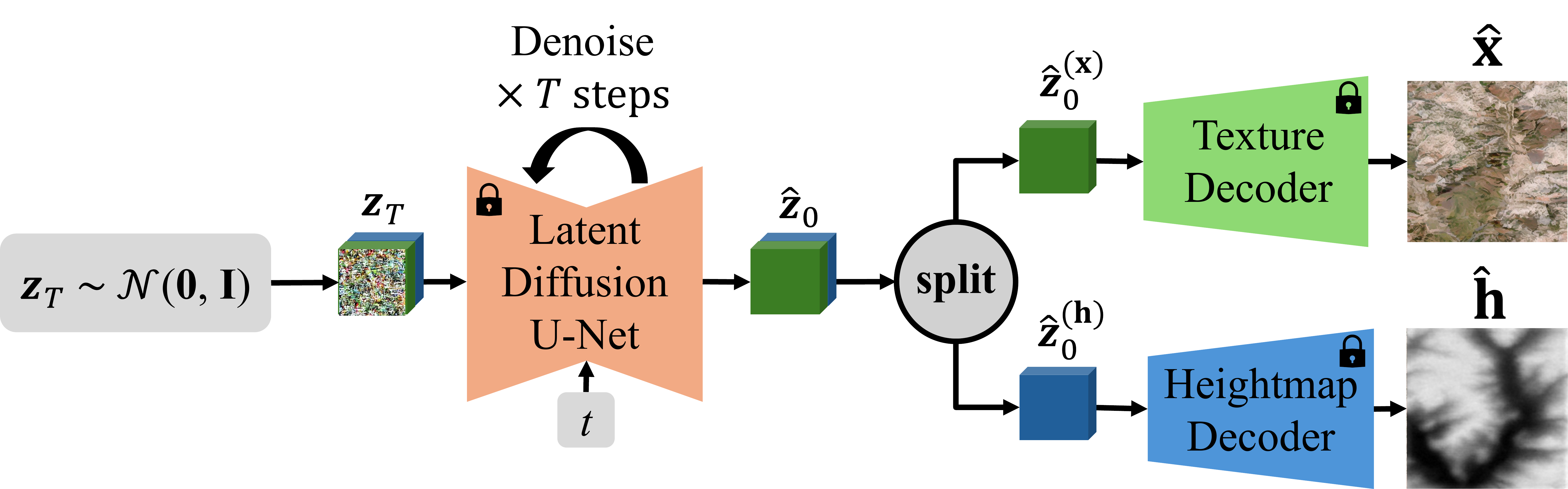}
  \caption{
  Inference flow of the unconditional joint generative model.
  }
  \label{fig:inference-uncond-terrain-ldm}
\end{figure}

\paragraph{
Inference.}
The inference process is illustrated in Figure~\ref{fig:inference-uncond-terrain-ldm}.
During inference, randomly sampled latent representations $\mathbf{z}^{(\mathbf{h})}_T$ and $\mathbf{z}^{(\mathbf{x})}_T$ are used as inputs, and noise is iteratively predicted and removed using the U-Net.
The resulting latents, $\hat{\mathbf{z}}^{(\mathbf{h})}_0$ and $\hat{\mathbf{z}}^{(\mathbf{x})}_0$, are then decoded into the final heightmap $\hat{\mathbf{h}}$ and texture image $\hat{\mathbf{x}}$ as follows:
\begin{align}
  \hat{\mathbf{h}} = \mathcal{D}_{h}(\hat{\mathbf{z}}^{(\mathbf{h})}_0), \quad \hat{\mathbf{x}} = \mathcal{D}_x(\hat{\mathbf{z}}^{(\mathbf{x})}_0),
  \label{eq:terrain-ldm-decode}
\end{align}
where $\mathcal{D}_h$ is the decoder for heightmaps, and $\mathcal{D}_x$ is the decoder from Stable Diffusion.

\paragraph{
Latent code fusion.
}
We extend the U-Net architecture used in LDM~\cite{rombach2022high} to jointly process textures and heightmaps in a unified latent space. Specifically, latent representations $\mathbf{z}^{(\mathbf{h})}_t$ and $\mathbf{z}^{(\mathbf{x})}_t$ are concatenated along the channel dimension to form a single input $\mathbf{z}_t = \text{cat}(\mathbf{z}^{(\mathbf{h})}_t, \mathbf{z}^{(\mathbf{x})}_t)$\ykA{, where $\text{cat}(\cdot,\cdot)$ denotes concatenation along the dimension of channels}.
The output is similarly defined as $\hat{\mathbf{z}}_0 = \text{cat}(\hat{\mathbf{z}}^{(\mathbf{h})}_0, \hat{\mathbf{z}}^{(\mathbf{x})}_0)$.
To support this structure, we double the number of input and output channels in the U-Net relative to the original LDM implementation.

\paragraph{
Leveraging Stable Diffusion prior.
}
\label{sec:fine-tuning-sd}
To enhance the quality of the generated terrain data, we leverage the prior knowledge of Stable Diffusion~\cite{rombach2022high}, which was pre-trained on large-scale image datasets.
We first extend the input and output layers of Stable Diffusion's U-Net to double the number of channels, aligning with the requirements of our model.
The additional channels are initialized with random weights.
We then fine-tune the extended U-Net using the loss function defined in Equation~(\ref{eq:terrain-ldm-loss}).
As Stable Diffusion is a text-conditioned model, we use a fixed prompt, ``\textit{A satellite terrain image.}'', during both training and inference.

\subsection{
Conditional Joint Generative Model}
\label{sec:controlnet}
To enable user control over the unconditional joint generation model described in Section~\ref{sec:fine-tuning-sd}, we extend it into a conditional model using ControlNet~\cite{zhang2023adding}.
Among various possible conditioning inputs, we follow the approach of Lochner et al.~\cite{lochner2023interactive} and use user-provided sketches representing terrain geometry (see Appendix~\ref{appendix_sketch} for \ykA{the} details).

\begin{figure}[t]
  \centering
  \includegraphics[width=\linewidth]{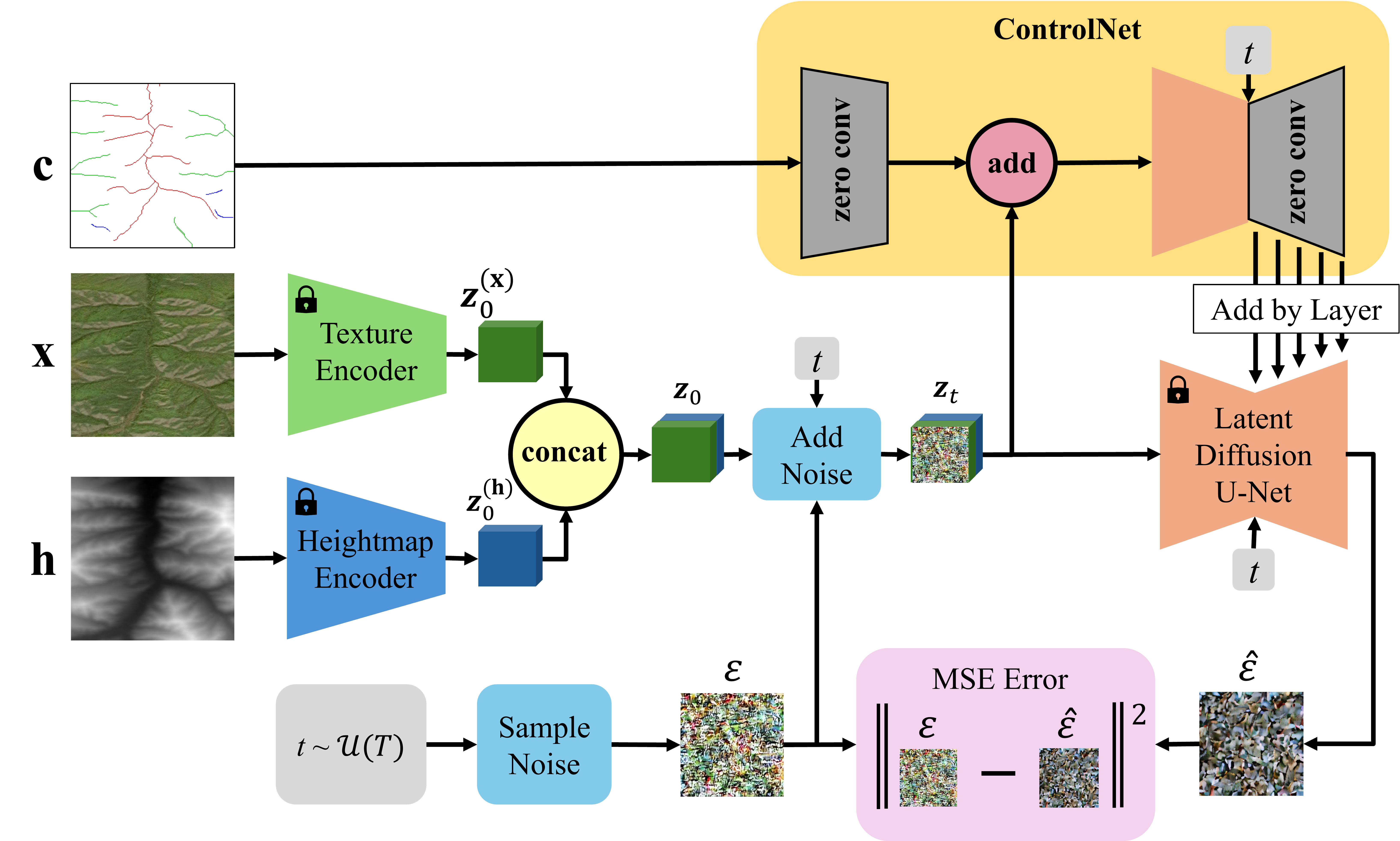}
  \caption{
  Training flow of the conditional joint generative model.
  }
  \label{fig:train-controlnet}
\end{figure}

Figure~\ref{fig:train-controlnet} provides an overview of the training process.
Let $\mathbf{c}$ be the input sketch image, and let $\epsilon_{\theta_c}$ denote the U-Net with ControlNet, parameterized by $\theta_c$.
The training loss $\ykA{\mathcal{L}}_c$ is defined as follows:
\begin{align}
  \ykA{\mathcal{L}}_c := \mathbb{E}_{\mathbf{z}^{(\mathbf{h})}_0, \mathbf{z}^{(\mathbf{x})}_0, \mathbf{c}, \mathbf{\epsilon} \sim \mathcal{N}(\mathbf{0}, \mathbf{I}), t} \left[ \left| \mathbf{\epsilon} - \mathbf{\epsilon}_{\theta_c}(\mathbf{z}^{(\mathbf{h})}_t, \mathbf{z}^{(\mathbf{x})}_t, \mathbf{c}, t) \right|^2 \right].
  \label{eq:controlnet-loss}
\end{align}
The inference process follows the same procedure as the unconditional model, except that the sketch input is additionally provided to the U-Net via the trained adapter.

\section{
Experiments}
\label{chap:experiments}

\paragraph{
Experimental settings.}

Our method was implemented using Python, PyTorch, and the Diffusers library~\cite{von-platen-etal-2022-diffusers}.
For the unconditional joint generative model, we used the AdamW optimizer~\cite{loshchilov2017decoupled} with a learning rate of $1 \times 10^{-4}$.
The noise scheduler was set to DDPM~\cite{ho2020denoising} with 1000 timesteps.
Training was performed for 200 epochs on an NVIDIA RTX A6000 GPU, taking approximately two days.
During inference, the reverse diffusion process also used 1000 steps and required about 75 seconds to generate a single $512 \times 512$ heightmap and texture pair.
For the conditional joint generative model, we used the Adam optimizer~\cite{kingma2014adam} with a learning rate of $1 \times 10^{-5}$, along with the same DDPM scheduler.
Training was conducted for 500 epochs on an NVIDIA RTX 6000 Ada GPU and took approximately seven days.
At inference time, the number of reverse diffusion steps was reduced to 20, enabling generation of a single $512 \times 512$ sample in approximately 2 seconds.

\paragraph{
Compared methods.}

As baselines for comparison, we trained two-stage generation models in which heightmaps and textures are generated separately using individual LDMs.
Specifically, we prepared two model pairs: one that first generates heightmaps independently and then generates textures conditioned on the heightmaps (i.e., $\mathbf{h} \rightarrow \mathbf{x}$), and another that first generates textures independently and then generates heightmaps conditioned on the textures (i.e., $\mathbf{x} \rightarrow \mathbf{h}$).
We also \ykA{consider comparison with} a GAN-based \ykA{terrain generation method by Spick et al.~\cite{spick2019realistic}}. \ykA{Unfortunately,}
their code and pre-trained model are not publicly available \ykA{and the official implementation of SGAN used in their method is too old to use}.
\ykA{We thus instead} trained PSGAN~\cite{bergmann2017learning}, an improved version of SGAN, using our dataset.

\subsection{
Quantitative Evaluation}

In this section, we quantitatively evaluate the quality of the generated textures, as well as the extent to which correlation between textures and heightmaps is preserved.
For this evaluation, we use 4,119 pairs of heightmaps and textures generated by each model, corresponding to the same number of data used during training.

\begin{table}[t]
  \centering
  \caption{
  Quantitative comparison of textures between the two-stage baselines and our method. Lower values indicate better performance. The best score is shown in \textbf{bold}, and the second-best is \underline{underlined}.
  }
  \begin{tabular}{lc}
  \hline
  Method                &  $\rm{FID_{CLIP}}$$\downarrow$\\ \hline \hline
  Baseline ($\mathbf{h} \rightarrow \mathbf{x}$)  & 24.0 \\ 
  Baseline ($\mathbf{x} \rightarrow \mathbf{h}$)  & 19.4 \\ 
  Ours (from scratch)                         & \underline{16.7}\\ 
  Ours (fine-tuning)                          & \textbf{9.8} \\  \hline
  \end{tabular}
  \label{tab:fid-comparison}
\end{table}

\begin{table}[t]
  \centering
  \caption{
  Quantitative comparison of textures between PSGAN~\cite{bergmann2017learning} and our method.
  }
  \begin{tabular}{lc}
  \hline
  Method                & $\rm{FID_{CLIP}}$$\downarrow$\\ \hline \hline
  PSGAN~\cite{bergmann2017learning}  & 22.1 \\ 
  Ours (from scratch)                         & \underline{16.7}\\ 
  Ours (fine-tuning)                          & \textbf{9.8} \\  \hline
  \end{tabular}
  \label{tab:fid-comparison2}
\end{table}

\begin{figure}[t]
  \centering
  \includegraphics[width=\linewidth]{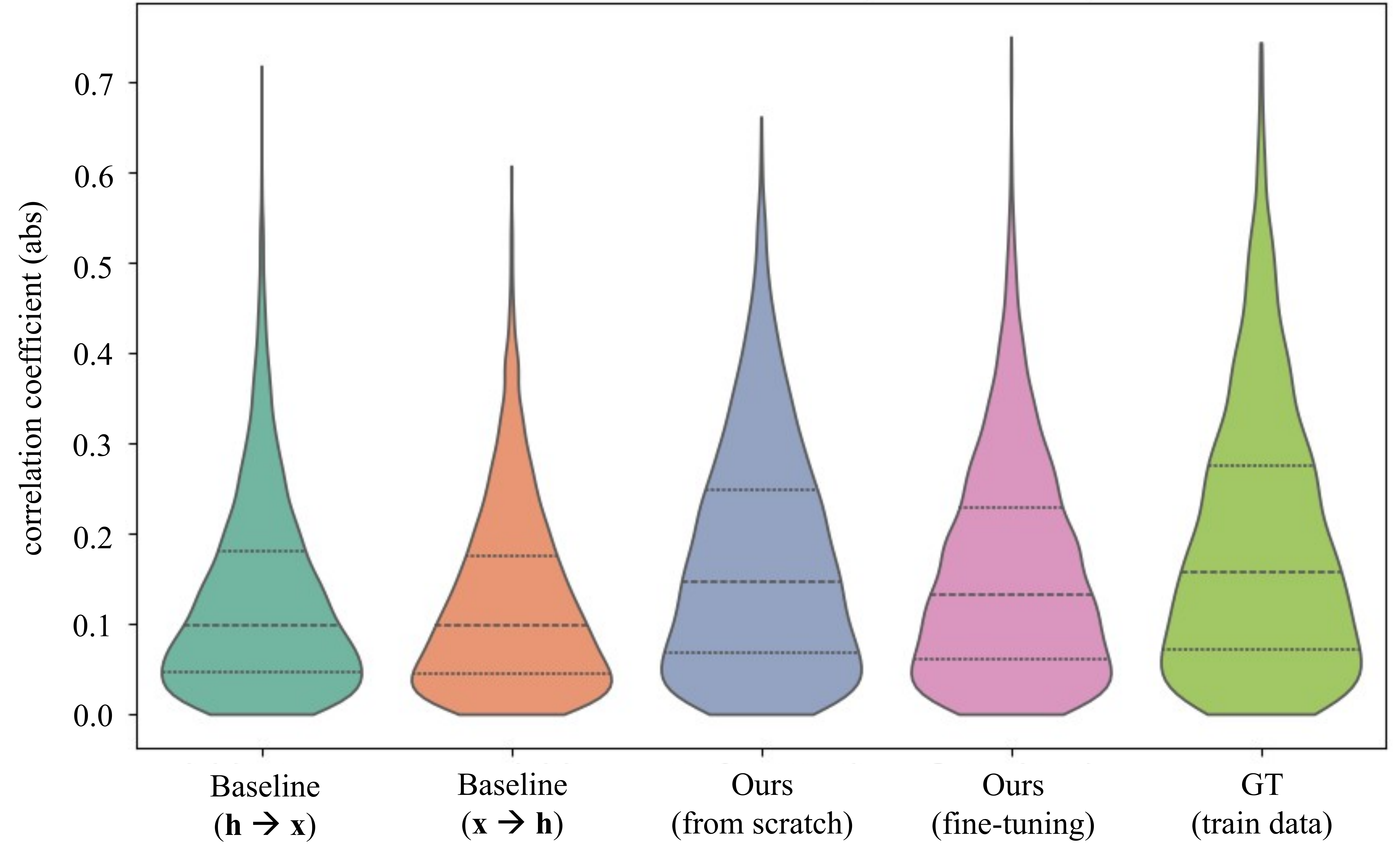}
  \caption{
  Comparison of correlation coefficients between textures and heightmaps generated by the two-stage baselines and our method. Dashed lines within each distribution indicate the first quartile, median, and third quartile.
  }
  \label{fig:correlation-comparison}
\end{figure}

\begin{figure}[t]
  \centering
  \includegraphics[width=\linewidth]{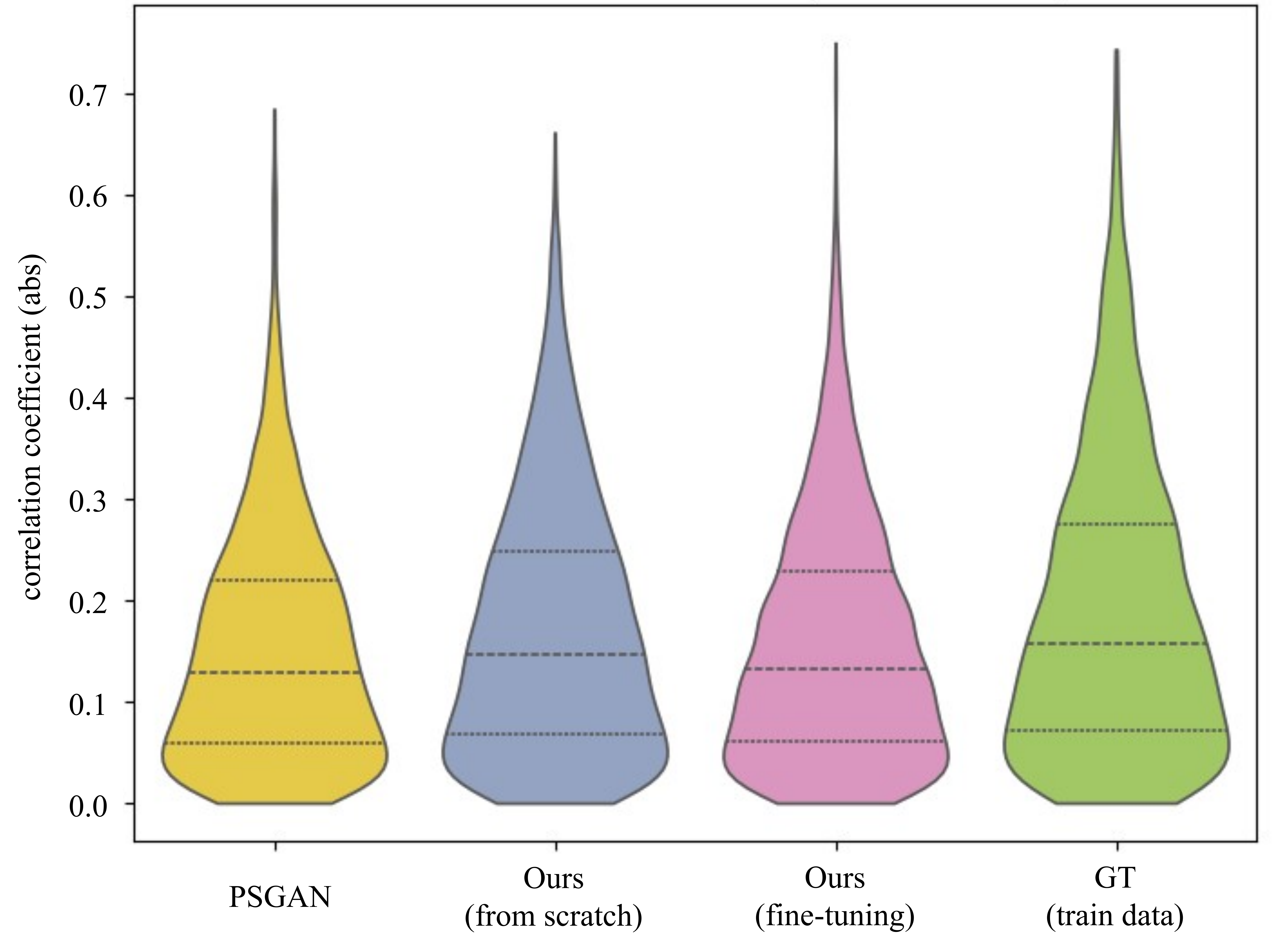}
  \caption{
  Comparison of correlation coefficients between textures and heightmaps generated by PSGAN~\cite{bergmann2017learning} and our method. 
  }
  \label{fig:correlation-comparison-2}
\end{figure}

\begin{table*}[t]
  \centering
  \caption{
  Statistics of correlation coefficients ($\uparrow$) between textures and heightmaps generated by the two-stage baselines and our method. IQR denotes the interquartile range. Values in parentheses indicate absolute differences from the ground-truth (training data). The best scores are shown in bold, and the second-best are underlined.}
  \label{tab:correlation-comparison}
  \begin{tabular}{lllllll}
    \toprule
    Method                                         & mean                    & std                  & 25\%                    & 50\%                    & 75\%                    & IQR                     \\
    \midrule
    Baseline ($\mathbf{h} \rightarrow \mathbf{x}$) & 0.13~(0.06)             & 0.11~(0.03)          & 0.05~(0.02)             & 0.10~(0.06)             & 0.18~(0.10)             & 0.13~(0.07)             \\
    Baseline ($\mathbf{x} \rightarrow \mathbf{h}$) & 0.12~(0.07)             & 0.10~(0.04)          & 0.04~(0.03)             & 0.10~(0.06)             & 0.18~(0.10)             & 0.13~(0.07)             \\
    Ours (from scratch)                            & \textbf{0.17~(0.02)}    & \textbf{0.12~(0.02)} & \textbf{0.07~(0.00)}    & \textbf{0.15~(0.01)}    & \textbf{0.25~(0.03)}    & \textbf{0.18~(0.02)}    \\
    Ours (fine-tuning)                             & \underline{0.16~(0.03)} & \textbf{0.12~(0.02)} & \underline{0.06~(0.01)} & \underline{0.13~(0.03)} & \underline{0.23~(0.05)} & \underline{0.17~(0.03)} \\
    GT (train data)                                & 0.19                    & 0.14                 & 0.07                    & 0.16                    & 0.28                    & 0.20                    \\
    \bottomrule
  \end{tabular}
\end{table*}

\begin{table*}[t]
  \centering
  \caption{Statistics of correlation coefficients ($\uparrow$) between textures and heightmaps generated by PSGAN [8] and our method. IQR denotes the interquartile range. Values in parentheses indicate the absolute differences from the GT (training data)}
  \label{tab:correlation-comparison-2}
  \begin{tabular}{lllllll}
    \toprule
    Method              & mean                    & std                  & 25\%                    & 50\%                    & 75\%                    & IQR                     \\
    \midrule
    PSGAN               & 0.15~(0.04)             & 0.11~(0.03)          & \underline{0.06~(0.01)} & \underline{0.13~(0.03)} & 0.22~(0.06)             & 0.16~(0.04)             \\
    Ours (from scratch) & \textbf{0.17~(0.02)}    & \textbf{0.12~(0.02)} & \textbf{0.07~(0.00)}    & \textbf{0.15~(0.01)}    & \textbf{0.25~(0.03)}    & \textbf{0.18~(0.02)}    \\
    Ours (fine-tuning)  & \underline{0.16~(0.03)} & \textbf{0.12~(0.02)} & \underline{0.06~(0.01)} & \underline{0.13~(0.03)} & \underline{0.23~(0.05)} & \underline{0.17~(0.03)} \\
    GT (train data)     & 0.19                    & 0.14                 & 0.07                    & 0.16                    & 0.28                    & 0.20                    \\
    \bottomrule
  \end{tabular}
\end{table*}

\begin{figure*}[t]
  \centering
  \includegraphics[width=\linewidth]{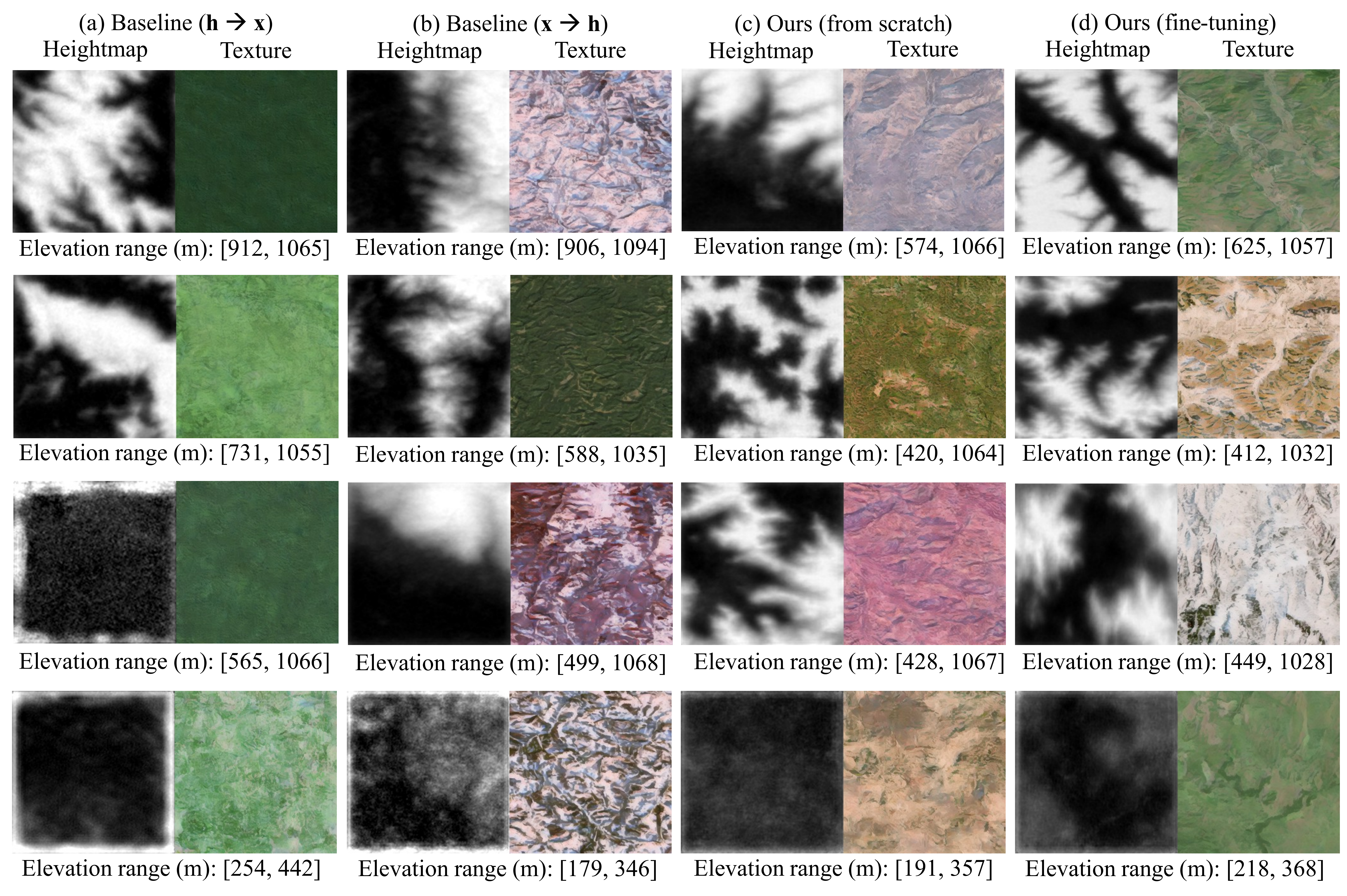}
  \caption{
  Comparison of terrain data generated by our method and the two-stage baselines.
  \ykA{H}eightmaps are normalized \ykA{within the range of [0, 1] for better display, where white and black represent higher and lower elevations, respectively. The original elevation range is shown} below each image.
  }
  \label{fig:generated-images}
\end{figure*}

\begin{figure*}[t]
  \centering
  \includegraphics[width=\linewidth]{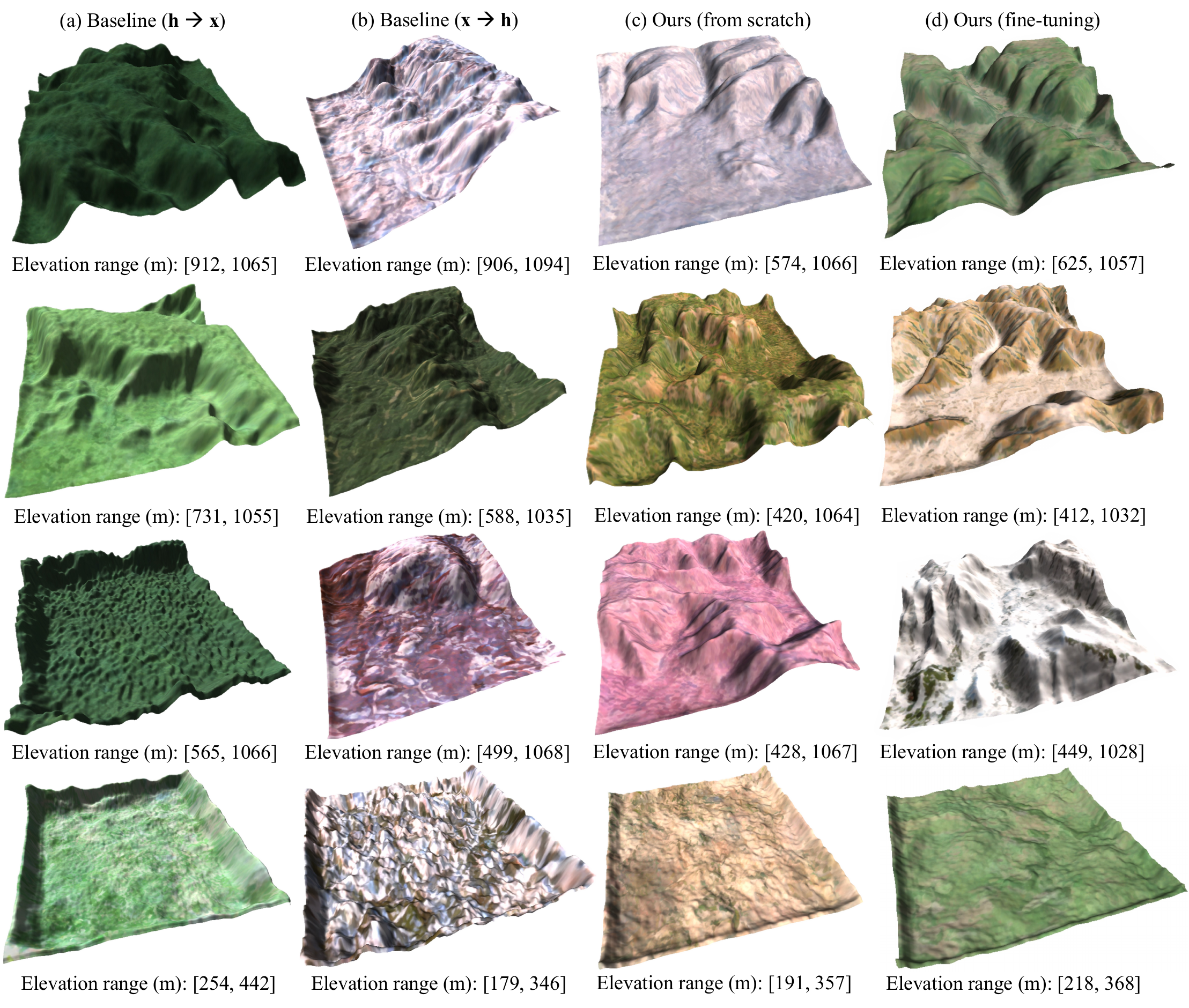}
  \caption{
  3D visualization of terrain data generated by our method and the two-stage baselines.
  }
  \label{fig:3d-visualization}
\end{figure*}

\begin{figure*}[t]
  \centering
  \includegraphics[width=\linewidth]{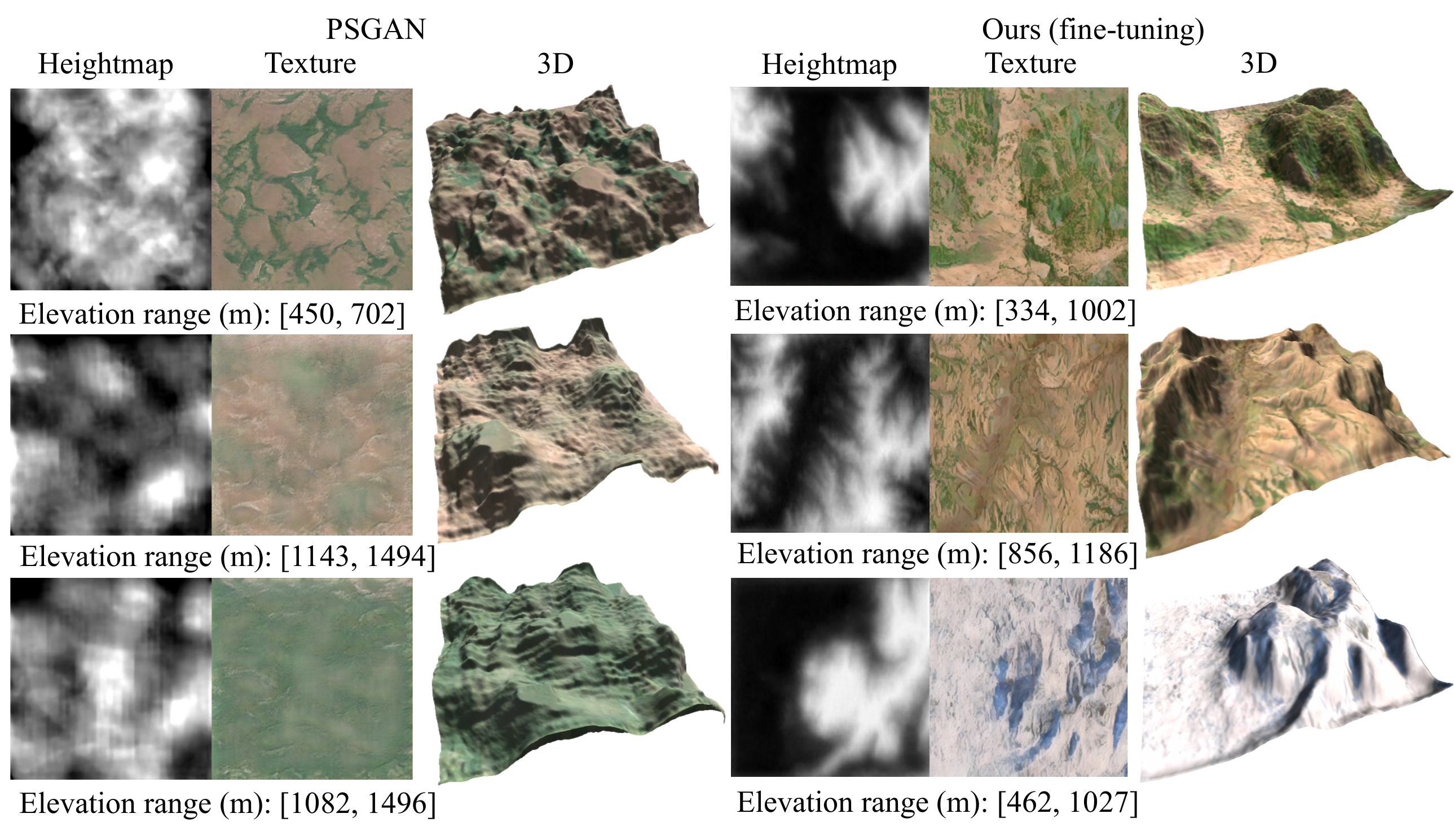}
  \caption{
  Comparison of terrain data generated by our method and PSGAN~\cite{bergmann2017learning}.
  }
  \label{fig:comparison-with-psgan}
\end{figure*}

\paragraph{
Texture quality. 
}

To evaluate the quality of generated terrain textures, we use CLIP Fréchet Inception Distance ($\rm{FID_{CLIP}}$)~\cite{Kynkaanniemi2022}, computed between the generated images and the training images.
While standard FID uses features from Inception-V3 trained on natural images, it may not reliably capture semantic similarity in terrain textures.
In contrast, $\rm{FID_{CLIP}}$ employs the CLIP visual encoder, which provides more semantically meaningful features and is thus better suited for texture quality evaluation.

Table~\ref{tab:fid-comparison} shows a comparison between our method and the two-stage baselines. Compared to the two-stage baselines, our joint generation model achieves lower scores, indicating higher similarity to the real textures.
Within our methods, the fine-tuned model outperforms the model trained from scratch. Table~\ref{tab:fid-comparison2} also demonstrates that our method outperforms the GAN-based method~\cite{bergmann2017learning}. 

\paragraph{
Correlation between textures and heightmaps.}

We quantitatively evaluate how well each model captures the correlation between heightmaps and textures.
Specifically, we compute the Pearson correlation coefficient between the pixel values of the heightmap and each channel of the corresponding texture image, \ykA{and} then average the results across channels.
The distribution of correlation coefficients across all generated pairs is visualized using violin plots for comparison.
The closer a model's distribution is to the training data, the better it reflects real-world terrain relationships.

The results are presented in Figure~\ref{fig:correlation-comparison}.
The two-stage generation models produce overall lower correlation values compared to the training data.
In contrast, our models yield distributions that more closely match the training distribution.
These results suggest that the joint generative models are more effective at capturing the correlation between heightmaps and textures. In addition, Figure~\ref{fig:correlation-comparison-2} shows a comparison between our method and PSGAN~\cite{bergmann2017learning}.
Although the correlation values of our method and PSGAN are comparable, PSGAN often yields artifacts in hightmaps, as discussed in Section~\ref{sec:quali_eval}.

 We also evaluate how well each model captures the correlation between heightmaps and textures. For each sample, we compute the Pearson correlation coefficient between the heightmap and each texture channel, and then average across channels. 
 The distributions of these correlation coefficients are visualized using violin plots. A distribution closer to that of the training data indicates better modeling of real-world terrain relationships.
Figure~\ref{fig:correlation-comparison} presents the results. Overall, the two-stage models produce lower correlation values compared to the training data, while our models yield distributions that more closely match the training distribution. Table~\ref{tab:correlation-comparison} summarizes the differences in correlation values between each method and the ground-truth data, further demonstrating the effectiveness of the joint generative models. In addition, Figure~\ref{fig:correlation-comparison-2} and Table~\ref{tab:correlation-comparison-2} show comparisons between our method and PSGAN~\cite{bergmann2017learning}. Although the correlation values of the two methods are comparable, PSGAN often produces artifacts in heightmaps, as discussed in Section~\ref{sec:quali_eval}.

\begin{figure*}[t]
  \centering
  \includegraphics[width=\linewidth]{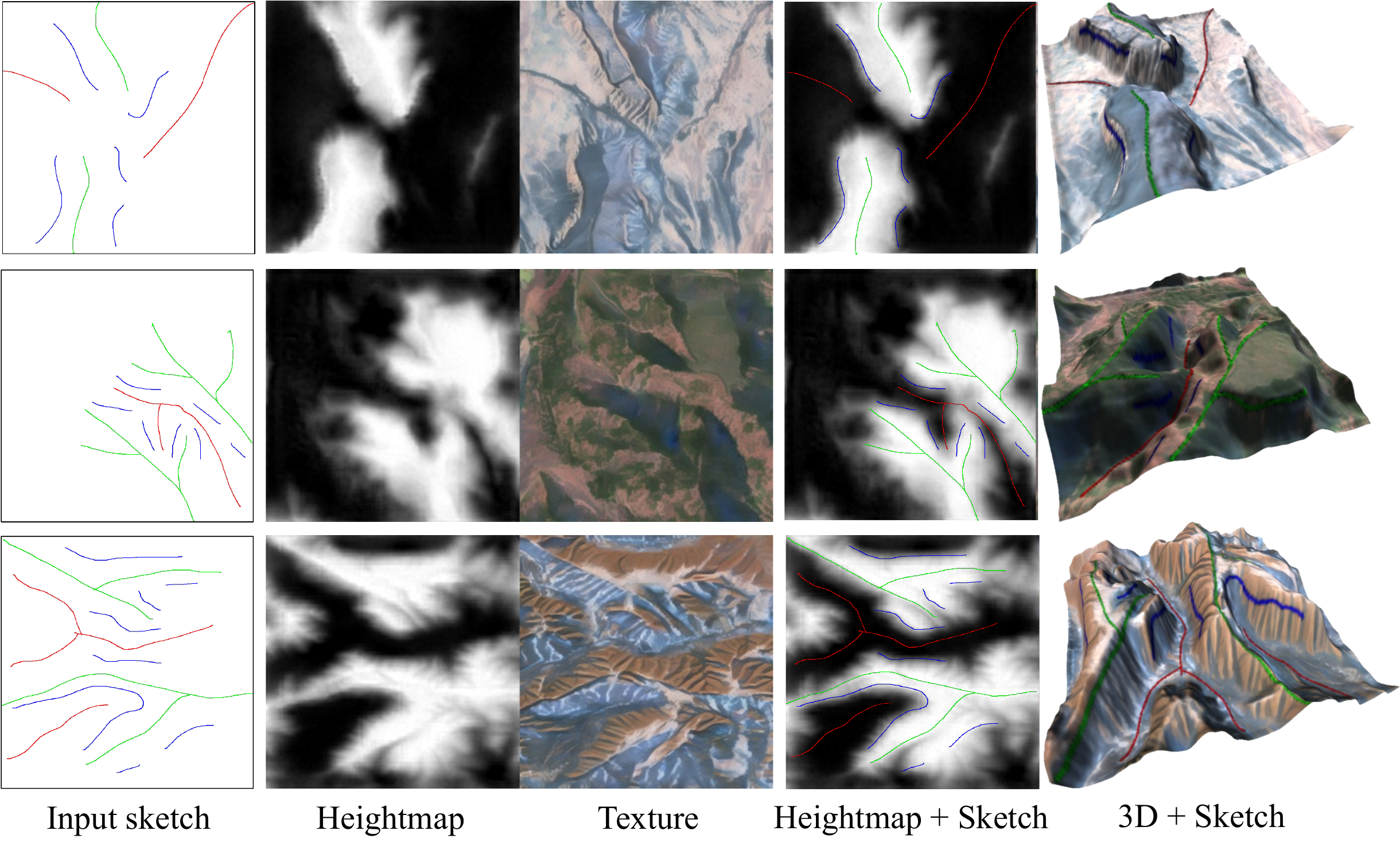}
  \caption{
  Results of terrain generation from sketch inputs. Red lines represent valleys, green lines indicate ridgelines, and blue lines denote cliffs.
  }
  \label{fig:sketch-conditioning}
\end{figure*}

\subsection{
Qualitative Evaluation
}\label{sec:quali_eval}
\paragraph{
Unconditional terrain generation.
}

Figure~\ref{fig:generated-images} shows heightmaps and textures generated by each model, and Figure~\ref{fig:3d-visualization} presents their corresponding 3D visualizations.
In Figures~\ref{fig:generated-images} and~\ref{fig:3d-visualization} (a), the two-stage model that generates textures conditioned on heightmaps produces textures with limited variation, often dominated by flat green tones, indicating low diversity.
In contrast, (b) shows that the model generating heightmaps conditioned on textures yields higher texture diversity, but frequently results in unrealistic color schemes.
The heightmaps also exhibit issues: many outputs contain unnatural structures such as elevated edges (e.g., the third and fourth rows in (a), and the fourth row in (b)).
Additionally, heightmap diversity is low. For example, in the first and third rows of Figure~\ref{fig:generated-images} (b), the outputs are often sharply divided into black and white regions.

Next, Figures~\ref{fig:generated-images} and~\ref{fig:3d-visualization} (c) show results from our model trained from scratch.
While some textures exhibit unnatural colors, both the heightmaps and textures demonstrate high diversity.
In (d), where our model is fine-tuned from Stable Diffusion~\cite{rombach2022high}, we observe similarly high diversity, along with more natural texture color schemes.
Compared to the outputs from (a), (b), and (c), the model in (d) produces fewer heightmaps with unnatural structures.
These results suggest that the model effectively leverages the prior knowledge of Stable Diffusion.
Notably, in the second row of (d), the texture appearance varies with elevation, indicating that the correlation between them has been successfully captured.

\paragraph{
  \higoA{Comparison with PSGAN.}
}

\higoA{Figure \ref{fig:comparison-with-psgan} shows a comparison of terrain data generated by our method and PSGAN~\ykB{\cite{bergmann2017learning}}.}
\ykB{The heightmaps generated by PSGAN exhibit noticeable noise in their elevation values despite we applied a median filter as a postprocessing, following Spick et al.~\cite{spick2019realistic}.}
\higoA{\ykB{Also, t}he heightmaps generated by PSGAN show repeated small-scale undulations, and the corresponding textures exhibit periodic patterns aligned with the elevation features.
This suggests that PSGAN learns to repeatively replicate local patterns from the training data.
As a result, PSGAN struggles to generate globally coherent terrain structures, such as continuous ridgelines or expansive plains.}

\higoA{In contrast, the heightmaps generated by our method exhibit broad ridgelines near the centers of mountainous regions, with finer ridgelines radiating outward.
The plains extending from the base of the mountains also appear smooth and natural, without any unnatural undulations.
These observations indicate that, compared to PSGAN, our method is more effective in generating terrain that reflects natural structure and realism.
}

\paragraph{
Conditional terrain generation. 
}

\begin{figure*}[t]
  \centering
  \includegraphics[width=\linewidth]{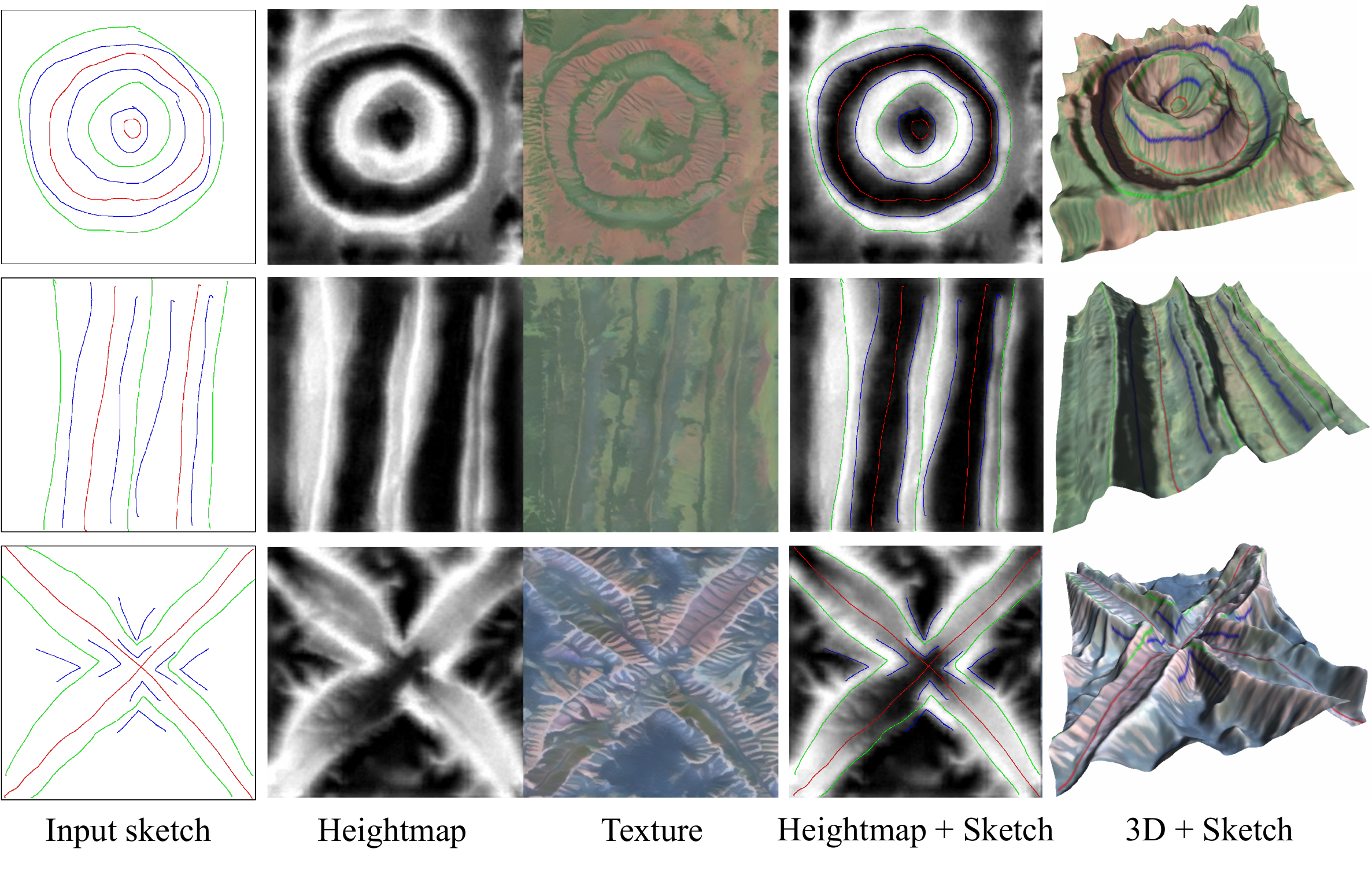}
  \caption{
  Results of terrain generation from sketch inputs representing extreme terrain geometry not found in the real world.
  }
  \label{fig:extreme-sketch-conditioning}
\end{figure*}

\begin{figure*}[t]
  \centering
  \includegraphics[width=\linewidth]{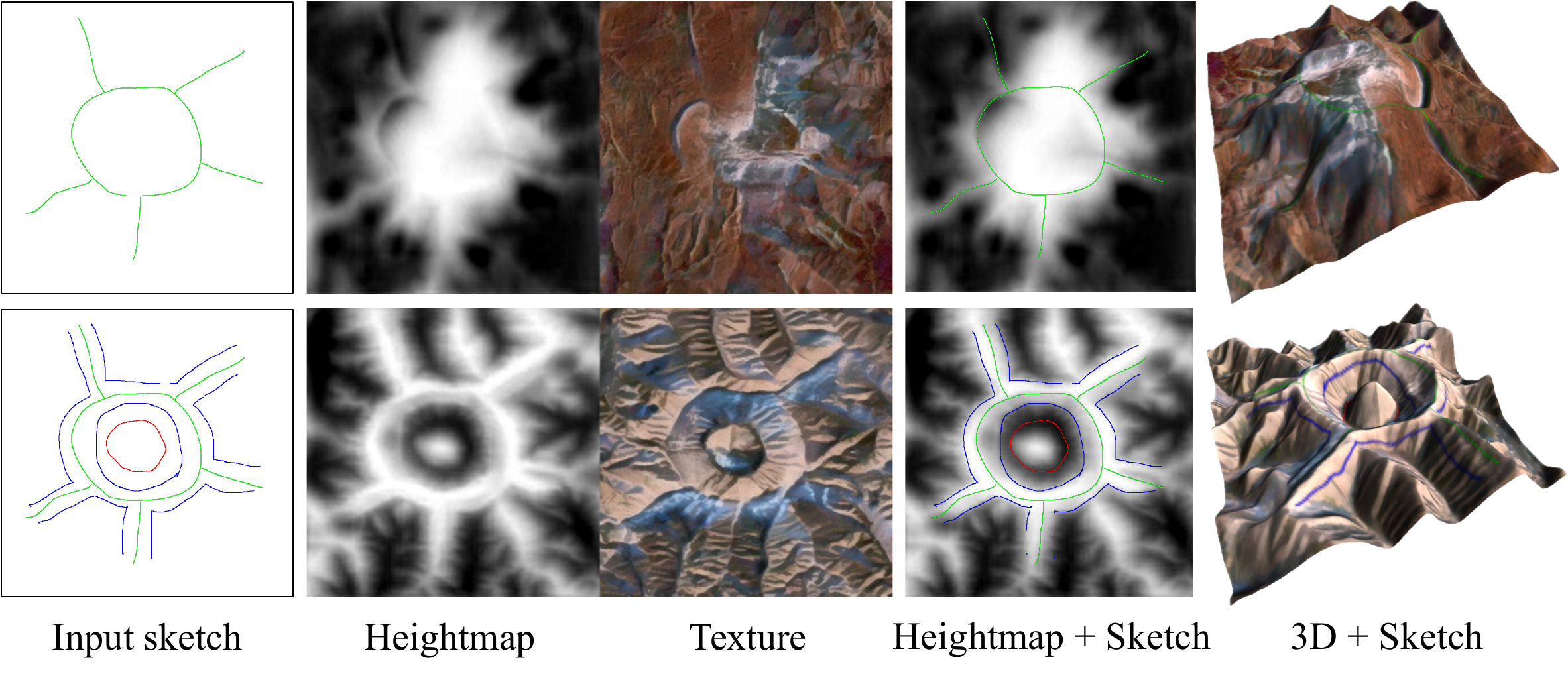}
  \caption{
  Results of conditional generation using sketches with similar overall shapes but varying levels of detail.
  }
  \label{fig:similar-sketch-comparison}
\end{figure*}

The results of conditional generation using user sketches are shown in Figure~\ref{fig:sketch-conditioning}.
The lines drawn in the input sketches align closely with elevation changes in the generated heightmaps.
As shown in Figure~\ref{fig:extreme-sketch-conditioning}, even sketches representing extreme terrain geometry not found in the real world are faithfully reflected in the outputs.
Figure~\ref{fig:similar-sketch-comparison} presents results from sketches with similar overall shapes but differing levels of detail.
Between two sketches sharing the same ridgeline, the one with more detailed surrounding features leads to outputs where the ridge is more clearly emphasized, resulting in a closer match to the input sketch.

\paragraph{
Texture Conditioning.
}

\begin{figure*}[t]
  \centering
  \includegraphics[width=\linewidth]{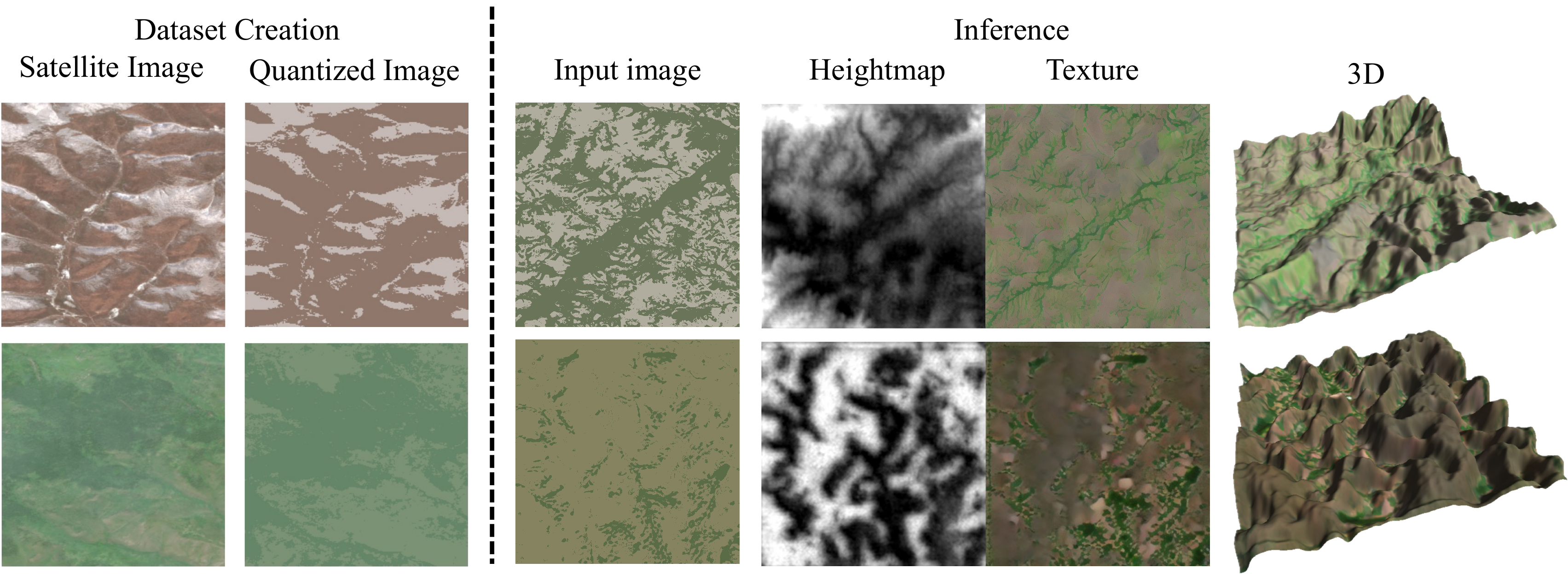}
  \caption{
  Details on our experiment on texture conditioning using ControlNet.
  }
  \label{fig:texture-conditioning}
\end{figure*}

To enable more flexible terrain generation, we explored texture control.
As shown on the left side of Figure~\ref{fig:texture-conditioning}, we first created a two-color-quantized texture dataset by approximating each texture image in the satellite image dataset with two representative colors. 
ControlNet was then trained using this dataset, following the same procedure described in Figure~\ref{fig:train-controlnet}. 
Given a user-drawn two-color image, the model synthesizes a texture conditioned on that image. Representative results are shown on the right side of Figure~\ref{fig:texture-conditioning}. These textures reflect the user-drawn inputs; however, the current model struggles to capture global terrain features, indicating room for improvement. Producing terrain-like outputs also requires complex input cues, and enabling simpler inputs remains an open challenge.

\section{
Discussion
}

We demonstrate that our method outperforms GAN-based approaches~\cite{bergmann2017learning, spick2019realistic}, which represent the state of the art in unconditional terrain generation. A recent concurrent study, MESA~\cite{mesa2025}, employs a diffusion-based model for terrain synthesis, as discussed in Section~\ref{sec:related-work}. However, MESA focuses on text-driven terrain generation, which differs from the unconditional and sketch-based tasks addressed in our work, making direct comparison beyond the scope of this study. We regard our method as complementary, and extending it to support textual input remains an interesting direction for future research.

\paragraph{
Evaluation by professional artists.
}

 We asked professional artists at the game studio Polyphony Digital Inc. to review our generated terrains and provide feedback. They noted that the textures vary plausibly with local slope and elevation, suggesting that our framework successfully captures essential geometry–appearance correlations. However, they also identified several limitations. Some textures contain baked-in shading, indicating the need for shading-free (albedo) textures via de-lighting. In addition, low-frequency noise sometimes causes blurring in the textures.
In terms of geometry, the relatively low resolution of both heightmaps and textures limited the evaluation of fine-scale realism relative to real-world terrains. The generated results also lacked key geomorphological features such as rivers, lakes, tree lines, and snow cover, which are important for producing convincing landscapes. The artists further emphasized the importance of explicitly defining spatial scale (e.g., $500~m \times 500~m$ versus $10~km\times10~km$), because recognizable landforms depend on scale.
For more in-depth evaluation, they suggested focusing on specific geographic regions to simplify the problem, improve output quality, and assess results in the context of region-specific landforms.

\paragraph{
Resolution
}
Resolution. Although our current experiments are conducted at a resolution of 512×512, scaling diffusion-based terrain generation to higher resolutions remains a significant challenge. While GANs are often considered efficient for high-resolution synthesis, recent advances in latent diffusion models (LDMs), such as SDXL~\cite{conf/iclr/2024}, demonstrate comparable capabilities. We believe our framework can be extended to higher resolutions by leveraging such improved LDMs, as well as approaches like latent-space super-resolution~\cite{jeong2025latent}, which support coherent upsampling in the latent domain and help preserve structural fidelity and fine detail. Furthermore, methods such as consistency models~\cite{song2023consistency}, which reduce the number of denoising steps, may lower computational costs during high-resolution inference. Exploring these approaches is an important direction for future work.

\section{
Conclusion
}

In this paper, we have proposed a method for jointly generating terrain heightmaps and textures.
By extending the latent space of an LDM, we model the joint distribution of heightmaps and textures, enabling unconditional terrain generation.
Furthermore, by integrating this model with ControlNet, we enable user control of terrain geometry through sketch input.
Qualitative results show that our method outperforms two-stage generation approaches and a GAN-based method.
Quantitatively, we confirmed that it captures a strong correlation between heightmaps and textures.
In the sketch-based control setting, our method successfully generates not only realistic terrain geometry but also extreme, user-defined geometries.

\paragraph{
Limitations and future work.
}

Although fine-tuning Stable Diffusion improved output quality in our method, artifacts in the generated textures remain.
Additionally, some heightmaps exhibit unnatural structures, suggesting that the overall quality is not yet sufficient for practical use in real-world 3D terrain modeling.
To address these limitations, we plan to expand the dataset and explore alternative data representations to further enhance output quality.
Beyond sketch-based input, incorporating other forms of conditioning, such as text prompts or semantic masks, could enable more precise control over texture colors and patterns.
These enhancements would significantly improve the practical utility of our approach as a terrain generation tool.

{\small
\bibliographystyle{ieeenat_fullname}
\bibliography{11_references}
}

\ifarxiv \clearpage \appendix \section*{Appendix}
\section{
Detail of Terrain Dataset
}
\label{appendix_dataset}
For texture acquisition, we targeted land areas between 60\textdegree S and 60\textdegree N observable by Sentinel-2.
We randomly sampled 500 locations, each covering a $1\tcdegree \times 1\tcdegree$ region in latitude and longitude.
To focus on natural terrain, we excluded regions with an average elevation below 100 meters, which are likely to include urban or artificial structures.
We further filtered locations using the Global Human Modification dataset~\cite{kennedy2019managing}, which quantifies human impact on the environment (spatial resolution: 1000~m; value range: [0, 1], with higher values indicating greater modification).
Regions with a human modification index of 0.3 or higher were excluded.
To mitigate climate-related sampling bias, we applied constraints based on the Köppen climate classification.
We excluded arid zones and focused on four climate categories: tropical, temperate, subarctic, and polar.
Sampling was limited to a maximum of 125 locations per category to maintain climate balance.
We did not enforce uniform sampling (e.g., 100 per category), as some climate zones contained fewer than 100 suitable regions.
After sampling, we applied additional filtering using the ESA WorldCover dataset~\cite{zanaga2021esaworldcover} (spatial resolution: 10~m).
We removed images including areas labeled as cropland, built-up regions, or water bodies, as were regions obscured by clouds or cloud shadows.
As a result, we obtained usable data from 276 locations.

\begin{figure}[h]
  \centering
  \includegraphics[width=\linewidth]{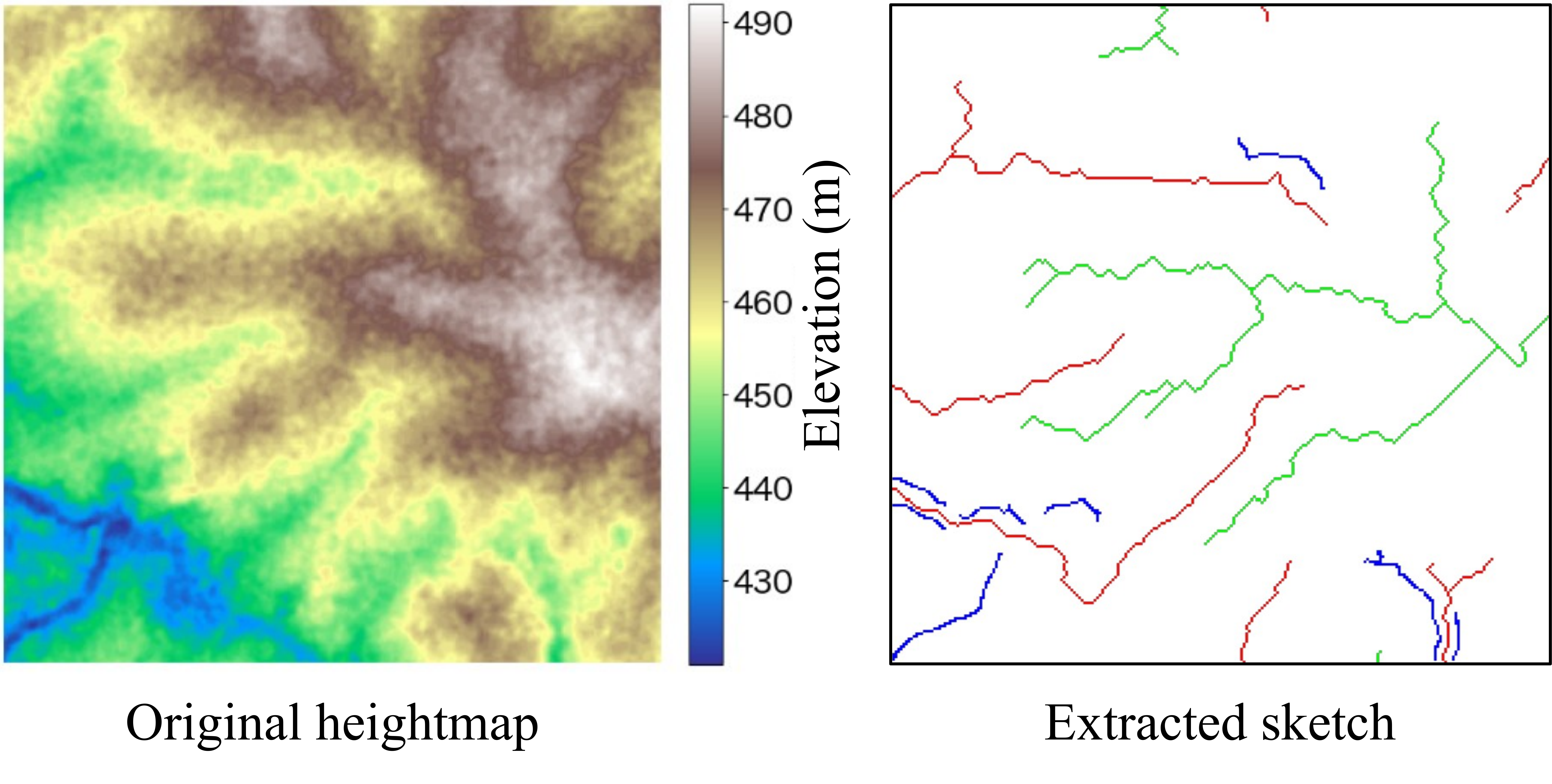}
  \caption{
  Sketch extraction (right) from a heightmap (left). The original heightmap is a grayscale image, but is visualized here with color mapping to emphasize elevation differences. Valleys are shown in red, ridgelines in green, and cliffs in blue.
  }
  \label{fig:sketch-data-creation}
\end{figure}

\section{
Detail of Sketch Dataset
}
\label{appendix_sketch}

As illustrated in Figure~\ref{fig:sketch-data-creation}, the sketches used in our method follow the method proposed by Lochner et al.~\cite{lochner2023interactive}, where valleys, ridgelines, and cliffs are represented by red, green, and blue lines, respectively.
To generate these sketches, we use the \texttt{FillDepressions} and \texttt{FlowAccumulation} functions with the \texttt{D8} method from the Python library RichDEM~\cite{RichDEM}, along with the Canny edge detection algorithm from the scikit-image library~\cite{van2014scikit}.
The \texttt{FillDepressions} function preprocesses the heightmap by filling sink areas, which is essential for simulating hydrological flow.
The \texttt{FlowAccumulation} function with the \texttt{D8} method assumes that water flows from each pixel to its steepest downward neighbor and computes the accumulated flow per pixel.
Pixels with flow values exceeding a predefined threshold are extracted as valleys.
To detect ridgelines, we invert the elevation values of the heightmap and apply the same flow accumulation process.
Finally, cliffs are extracted by applying the Canny edge detection algorithm to the heightmap.

 \fi

\end{document}